# A Novel VLSI Technology to Manufacture High-Density Thermoelectric Cooling Devices

Howard Chen, Louis Hsu, and Xiaojin Wei
IBM Corporation, New York, U.S.A.

***Abstract*** - This paper describes a novel integrated circuit technology to manufacture high-density thermoelectric devices on a semiconductor wafer. With no moving parts, a thermoelectric cooler operates quietly, allows cooling below ambient temperature, and may be used for temperature control or heating if the direction of current flow is reversed. By using a monolithic process to increase the number of thermoelectric couples, the proposed solid-state cooling technology can be combined with traditional air cooling, liquid cooling, and phase-change cooling to yield greater heat flux and provide better cooling capability.

## 1. Traditional Thermal Technology

The International Technology Roadmap for Semiconductors predicts that power dissipation and thermal problems are major design challenges to continued scaling. To meet the power demand of each application, today's prevailing cooling technologies can be divided into 4 categories: natural air convection, forced air convection, fluid phase change, and liquid cooling. For example, heat sinks have been widely used as air cooling devices through either passive natural convection or forced convection when the airflow is induced by a fan. For low-power system level cooling, stamped heat sinks provide an economic solution by stamping copper or aluminum sheet metal into desired shapes. For high-power system level cooling, extruded heat sinks allow greater heat dissipation through increased surface area of complex fin structures. Under forced convection, the thermal efficiency of high fin density heat sinks can be further improved with the addition of fans that increase the airflow [1].

For high-end cooling applications, two-phase heat transfer mechanism within a heat pipe provides a highly effective thermal solution. As the heat enters the evaporator section of a heat pipe, it causes the working fluid to vaporize. The vaporized fluid creates a pressure gradient that forces the vapor to flow along an adiabatic pipe to the cooler condenser section. As the vapor condenses, its latent heat of vaporization is released. The condensed working fluid is then returned to the evaporator by the capillary force of the wick structure embedded inside the heat pipe [2].

After a decade's absence, IBM announced its return to water cooling in 2005 [3]. A typical liquid cooling system uses a pump to circulate a single-phase thermally conductive liquid to remove the excess heat from the processor and release it to the ambient air flowing through a liquid-to-air heat exchanger. However, as the power density continues to increase, alternative cooling technologies such as thermoelectric cooling are being explored to meet the thermal design challenges of the future. It has been shown that when combined with two-phase passive cooling devices, thermoelectric devices could further improve the thermal efficiency beyond the cooling capability of conventional technologies [4].

## 2. The Principles of Thermoelectric Cooling

A thermoelectric module is a solid-state heat pump that consists of multiple pairs of n-type and p-type thermoelectric elements that are connected electrically in series and thermally in parallel (Fig. 1). Since the n-type thermoelectric material is doped with an excess of electrons and the p-type thermoelectric material is doped with an excess of holes, they can be used to conduct electricity and heat in a thermoelectric module. By applying a direct current (DC) through the n-type and p-type thermoelectric elements, the electric current will flow alternately through each n-type device and p-type device, and travel back and forth between the cold plate (chip side) and the hot plate (heat sink side) of a thermoelectric module. As the electrons in n-type devices and the holes in p-type devices move from the cold plate to the hot plate, they carry heat from one side of the thermoelectric module to the other side of the thermoelectric module.

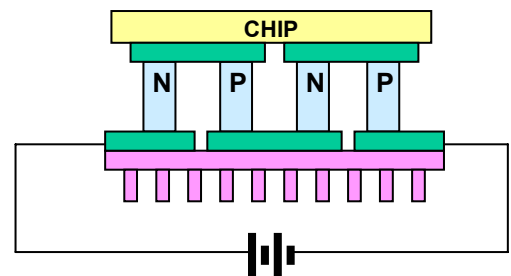

Figure 1. Thermoelectric cooling module.

The principles of heat transfer in a thermoelectric module can be explained by the Peltier effect [5]. When the electrons and holes move from one junction (cold plate) to the other junction (hot plate), the heat is absorbed at the cold plate in order to move additional electrons from the valence band to the conduction band, so that additional pairs of electrons and holes can be generated to replenish and compensate for the loss of charged carriers. The heat is released at the other junction (hot plate) as the electrons reunite with holes. Therefore, by applying a DC voltage to a thermoelectric module, heat can be moved from one side of the module to the other side of the module. Since the direction of the heat flow depends on the polarity of the DC voltage applied to the thermoelectric module, heat can be pumped through a thermoelectric module in either direction. Consequently, a thermoelectric module may be used for both heating and cooling, which makes it suitable for applications that require precise temperature control. Furthermore, as a result of the Seebeck effect [5], the thermoelectric module can be used for power generation, where a current is generated due to the temperature differential across the thermoelectric module.

Most thermoelectric modules have an equal number of n-type and p-type elements and each pair of n-type and p-type elements forms a thermoelectric couple. The thermoelectric couple elements and their electrical interconnects are usually mounted between two thermally conductive ceramic substrates, which hold the thermoelectric module together mechanically and insulate the thermoelectric elements electrically. The cooling capacity of a thermoelectric module is a function of the DC current applied and the temperature difference between the hot plate and the cold plate.

In order to properly adjust the voltage level of thermoelectric elements and the cooling capacity of the thermoelectric assembly, Chu [6] proposed a method to integrate the thermoelectric assembly with a programmable power control circuit in the thermal dissipation assembly of an electronic module. The control circuit therefore determines how much power is delivered to the thermoelectric assembly through conductive power planes in the supporting substrate on the backside of the chip.

The characteristics of the thermoelectric cooling module are light-weight, maintenance-free, and environmentally friendly. With no moving parts, it is also acoustically silent and electrically quiet. A thermoelectric module can provide active heating and cooling in the same device with precise temperature control (within 0.1 °C). It can also operate in a wide range of temperatures, from sub-ambient cooling to low temperature (-80 °C) cooling.

### 3. Thermoelectric Material

The ideal Peltier cooling rate of a thermoelectric module is equal to $Q_P = \pi \cdot I = \alpha \cdot T_c \cdot I$, where $\pi$ is the Peltier coefficient, $\alpha$ is the Seebeck coefficient, $T_c$ is the cold-plate temperature in degrees Kelvin, and I is the DC current. However, in reality, the heat will also be conducted from the hot plate back to the cold plate. The Fourier heat conduction rate is equal to $Q_C = \kappa \cdot \Delta T$, where $\kappa$ is thermal conductance and $\Delta T$ is the temperature difference between the hot plate and the cold plate. In addition, there will be Joule heating, which equals $\frac{1}{2} \cdot I \cdot R^2$, due to the electrical resistance R of thermoelectric elements. As a result, the net cooling rate of a thermoelectric module can be calculated from the following formula

$$Q_{NET} = \alpha \cdot T_c \cdot I - \kappa \cdot \Delta T - \textbf{½} \cdot I \cdot R^2 .$$

In order to increase the Peltier cooling rate and reduce the back flow of Fourier heat conduction from hot plate to cold plate, thermoelectric material must exhibit a high Seebeck coefficient, low thermal conductivity and low electrical resistance. For example, metal alloys are known to have high thermal and electrical conductivities, but since their Seebeck coefficients are inherently low, metals are not suitable for thermoelectric cooling. On the other hand, extrinsic alloy semiconductors are known to offer reasonable electric conductivity while maintaining low thermal conductivity. They also exhibit relatively large Seebeck coefficients and narrow band gap. These semiconductor alloys include zinc antimonite, indium antimonite, lead telluride, bismuth telluride, silicon germanium (SiGe), and Li-Ni-O compounds. Based on the cost and the compatibility of the deposition process with VLSI integration, SiGe is a suitable thermoelectric material for thin-film micro-fabrication. SiGe can be grown epitaxially, or deposited at low temperature (<300 °C) by using chemical vapor deposition. It can be doped with an ion implant process to form p-type and n-type thermoelectric couple elements, after dopants are activated via rapid thermal or laser annealing.

### 4. Processing Steps to Form Thermoelectric Array

The heat pumping capacity of a thermoelectric cooling module can be improved by increasing the number of thermoelectric couple elements and arrays. Therefore, the reduction of the feature sizes of thermoelectric couple elements using advanced thin-film material and technologies will lead to significant improvement in the heat pumping capacity of the thermoelectric cooling module. For example, the cooling power density of a thin film thermoelectric cooler with 20-um legs is about two orders of magnitude greater than that of a commercial thermoelectric cooler with 2-mm legs, because the cooling power density is inversely proportional to the length of the thermoelectric legs [7]. Using thin-film materials and a wafer scale process, a thermoelectric cooler could achieve a thickness of 1-2 um, single stage $\Delta T$ of 50-70 °C and a tunable performance of 10–1,000 W/cm$^2$ [8][9].

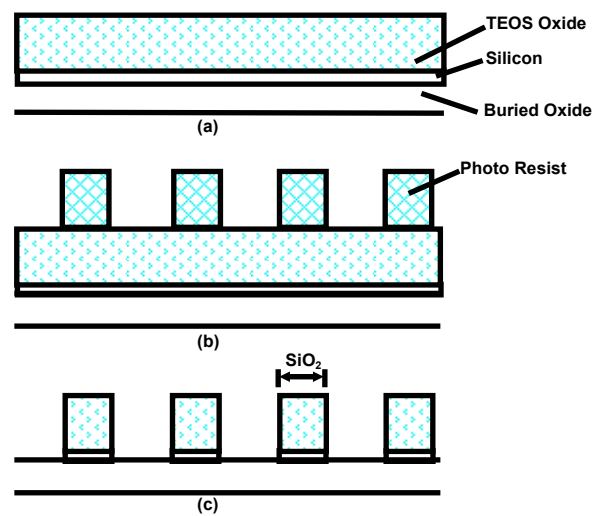

Figure 2. Wafer preparation to fabricate TE devices.

      

In this section, we propose a new technology to form thermoelectric (TE) couple elements on a bulk silicon or SOI substrate. For the purpose of illustration, an SOI wafer is shown in Fig. 2 to demonstrate the process of forming self-aligned thermoelectric couple elements and interconnects.

First, a layer of dielectric material ($SiO_2$) is deposited on the surface of an SOI wafer by using a plasma-enhanced tetra-ethyl-ortho-silicate (TEOS) deposition process (Fig. 2a). The thickness of the TEOS oxide layer will determine the height of the thermoelectric cooling couple elements. A photolithography process is then used to define the array patterns of thermoelectric couple elements (Fig. 2b), followed by an etching process to create the structural voids needed in the dielectric and the silicon layers to form arrays of thermoelectric couple elements (Fig. 2c).

Next, a chemical vapor deposition (CVD) process is used to deposit a second dielectric material such as nitride to fill the voids (Fig. 3a). The top surface of the CVD nitride is planarized and flushed with the top surface of the first dielectric material (TEOS oxide). The first dielectric material (TEOS oxide) is then selectively removed by using reactive ion etching (RIE) technique to create a three-dimensional structure where the thermoelectric cooling material can be deposited (Fig. 3b). The sidewalls of the second dielectric material (CVD nitride) can be tapered with a desired angle (between 75° and 90°), using controlled wet etching to facilitate the ensuing angle implantation (Fig. 3c).

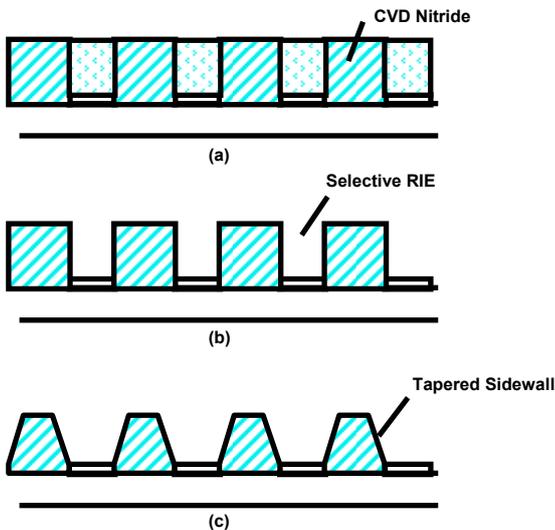

Figure 3. Reactive ion etching to remove TEOS oxide.

A thin film of thermoelectric material such as SiGe can now be deposited on the top surface or formed by using the exposed silicon surface as the seed for epitaxial growth. Since the layer of thermoelectric cooling material conforms with the contour of the tapered sidewalls of CVD nitride, angle implantation can be employed by doping one side of the thermoelectric material with an n-type dopant, and doping the other side of the thermoelectric material with a p-type dopant.

Following angle implantation, which has the advantage of not requiring any photo mask, the thermoelectric materials along the sidewalls should be properly annealed to form the n-type and p-type thermoelectric couple elements (Fig. 4a). The newly formed n-type and p-type thermoelectric elements on the sidewalls of CVD nitride are insulated and covered with nitride spacers (Fig. 4b). The exposed top and bottom parts of the SiGe epitaxial layer are converted into metal silicides by a self-aligned silicidation process. The top and bottom metal silicides are used as low-resistance contacts to electrically connect the adjacent p-type and n-type thermoelectric couple elements.

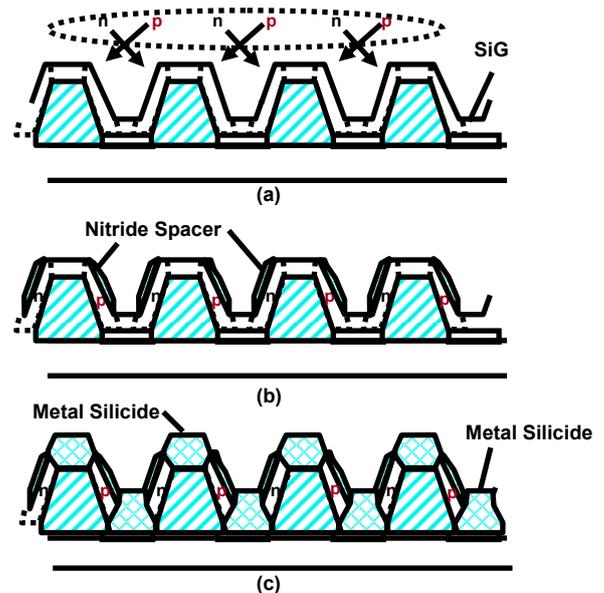

Figure 4. Angle implantation of n and p-type dopants.

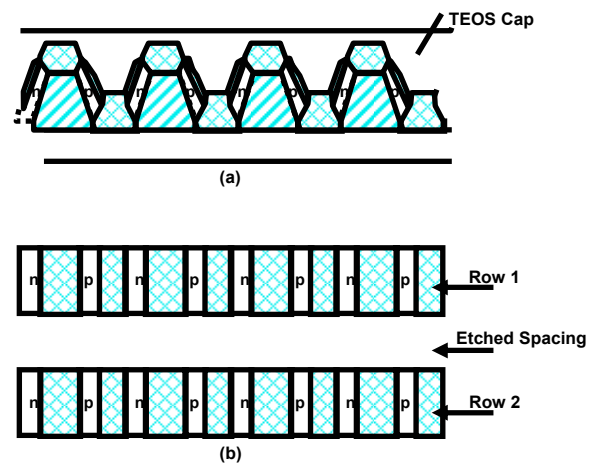

Figure 5. Formation of thermoelectric couple arrays.

After silicidation, a thick layer of insulating material can be deposited on the top of the structure by using TEOS-based $SiO_2$ to cap and protect the thermoelectric device (Fig. 5a). Array structures of the thermoelectric cooling couple elements can now be formed by a photolithographic patterning process, followed by an etching process to remove the metal silicides, SiGe thin films, and the sidewall spacers between adjacent rows of thermoelectric couple elements (Fig. 5b). As the etching stops at the buried oxide layer, arrays of high-density thermoelectric cooling elements are formed.

## 5. Integration of Thermoelectric Cooling Module

The use of a high-density thermoelectric module allows cooling to take place inside the chip. The integration of thermoelectric cooling devices on a chip, however, requires an effective means such as air cooling by heat sink or liquid cooling through micro-channels to remove the heat from thermoelectric couple elements. To facilitate various heat removal requirements, three different methods are proposed to integrate thin-film thermoelectric cooling elements on the chip. The first method fabricates thermoelectric cooling devices on the backside of a semiconductor chip. Since the thermoelectric elements can be formed at low temperatures (<300 °C), they should be fabricated after the integration and metallization of device circuits on the front side of the chip are complete and protected. After the front side of the chip is mounted on a carrier via solder balls, a heat-sink device can be attached on the backside of the chip, using a thermal interface material such as silicone-based greases, elastomeric pads, thermally conductive tapes, or thermally conductive adhesives. A wire-bonded cable is then used to supply the power to thermoelectric cooling devices.

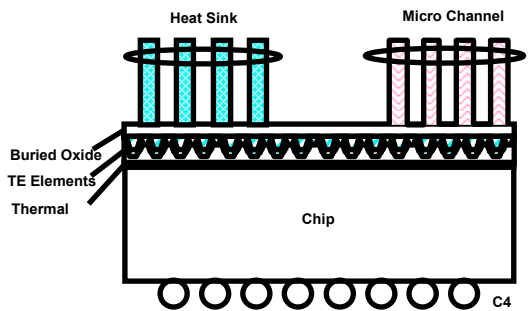

Figure 6. Backside bonding of a TE module.

Fig. 6 shows a second implementation, where an integrated thermoelectric cooling module is bonded to a chip. The integrated thermoelectric cooling module consists of arrays of thermoelectric couple elements fabricated separately on the front side of an SOI wafer, while the heat sink or micro-channel devices can be attached to the backside of the substrate. Deep trenches can be etched on the backside of the wafer, until the etching stops at the buried oxide layer, and filled with thermally conductive material such as metal, CVD diamond, or CVD alumina. After planarization and the removal of the original silicon substrate, the completely integrated thermoelectric cooling module can be glued to the backside of a semiconductor chip using thermal paste.

For the most critical applications, it may be possible to place the thermoelectric couple elements directly adjacent to the hot spots where the heat is generated. Since thermoelectric devices are capable of conducting heat in either direction, more sophisticated designs can also incorporate thermal sensors to precisely control the on-chip temperature by using the thermoelectric module as either a cooling or heating device.

In certain applications, the power dissipation may require multiple thermoelectric cooling modules to be stacked up to further improve the thermal efficiency. For instance, the theoretical maximum temperature difference $\Delta T$ between the cold side and the hot side of a typical single-stage thermoelectric cooler module is between 65° and 70°C [5]. If a higher $\Delta T$ is desired, then a multistage thermoelectric cooler could be implemented by stacking several single stage thermoelectric coolers on top of each other.

## 6. Thermal Simulation Results

A thermal resistance network similar to [10] is developed to model the TE structure in Fig. 6. The main components of the thermal resistance network include the bulk silicon resistance $\theta_{si}$, the thermal interface resistance $\theta_{TIM1}$ between the chip and the TE cooler, the thermal interface resistance $\theta_{TIM2}$ between the TE cooler and the external heat sink, and the resistance of the external heat sink $\theta_{ext}$ (Fig. 7). The thermoelectric cooler is characterized by its cold-side heat flow $Q_c$, hot-side heat flow $Q_h$, the cold-side temperature $T_c$, and hot-side temperature $T_h$. In addition, the input electrical power needed to provide thermoelectric cooling is represented by $Q_e$.

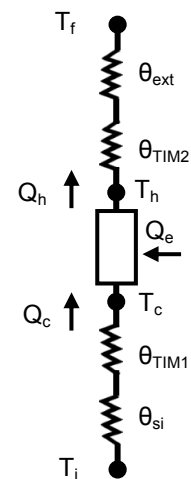

Figure 7. Thermal resistance of TEC structure.

From [11], the cooling capacity of the TE cooler is

$$Q_c = N\left(\alpha_{eff} T_c I - K_{eff}\left(T_h - T_c\right) - \frac{1}{2}I^2 R\right). \quad (1)$$

The required input electrical power is

$$Q_e = N\left(\alpha_{eff} I\left(T_h - T_c\right) + I^2 R\right). \quad (2)$$

To balance the energy from (1) and (2), the heat rejection on the hot side of the TEC can be derived as

$$Q_h = N\left(\alpha_{eff}T_h I - K_{eff}\left(T_h - T_c\right) + \frac{1}{2}I^2R\right). \quad (3)$$

Assuming one-dimensional heat transfer, the hot-plate temperature is

$$T_h = T_f + Q_h\left(\theta_{ext} + \theta_{TIM2}\right) \quad (4)$$ and the

cold-plate temperature is

$$T_c = T_j - Q_c\left(\theta_{si} + \theta_{TIM1}\right). \quad (5)$$

The parallel thermal conductance of the n-type and p-type TE elements is

$$K_{eff} = \left(\frac{k_n W_n \tau_n}{L_n} + \frac{k_p W_p \tau_p}{L_p}\right), \quad (6)$$

where the width W, length L, and pitch P of the TE elements are assumed to be 10 um and the thin-film thickness $\tau$ is assumed to be 2 um in our baseline simulation. Similarly, the serial electrical resistance of the n-type and p-type TE elements is given by

$$R = \left(\frac{\rho_n L_n}{W_n \tau_n} + \frac{\rho_p L_p}{W_p \tau_p}\right). \quad (7)$$

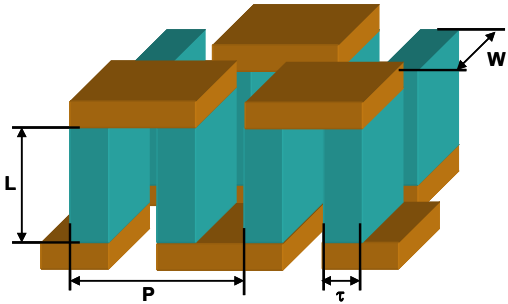

Figure 8. 3D structure of TEC elements.

A pictorial rendering of the TEC array without the structural dielectric film is shown in Fig. 8. For a given chip power input $Q_c$ and a TE cooler with known geometric parameters and thermo-electrical properties, the chip junction temperature $T_j$ can be derived from equations (1)-(7). For a chip without the TEC structure, $T_j$ is calculated by removing the TEC model and thermal resistance $\theta_{TIM2}$ from the resistance network in Fig. 7. The parameters used in our baseline thermal simulation are listed in Table 1.

Table 1. Baseline parameters for thermal simulation

| | |
|---|---|
| Effective Seebeck coefficient $|\alpha_n|+|\alpha_p|$ | 440 uV/°C |
| Electrical resistivity $\rho_n = \rho_p$ | 10 Ω·um |
| Thermal conductivity $\kappa_n$ | 1.70 W/m°C |
| Thermal conductivity $\kappa_p$ | 1.45 W/m°C |
| Thermal resistance $\theta_{ext}$, $\theta_{TIM1}$, $\theta_{TIM2}$ | 0.1 °C/W |
| TE element length L, width W, pitch P | 10 um |
| TE Thin film thickness $\tau$ | 2 um |
| Chip size | 1cm × 1cm |
| Chip power | 50 W |
| Coolant temperature $T_f$ | 20°C |

Fig. 9 shows that for a given power density (50 W/cm$^2$), an optimum current can be applied to maximize the benefit of the thermoelectric cooling. For instance, by applying a current of 4 mA to the TEC structure, the chip junction temperature can be lowered by an additional 20°C, as $T_j$ reaches 6.5°C below the ambient temperature $T_f$. In addition, Fig. 9 shows that as the input current increases, the coefficient of performance (COP) decreases. Therefore, although there are two current values that might deliver the same cooling effect, the lower current will always provide a better COP. Fig. 10 shows the TEC cooling effects for different power densities at their respective optimum current conditions. Since the efficiency of Peltier cooling is limited by the Fourier heat conduction and Joule heating, the benefit of using a TEC diminishes when the power density exceeds 120 W/cm$^2$.

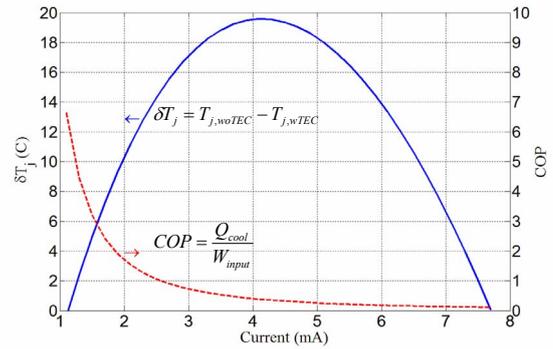

Figure 9. Optimal TEC current.

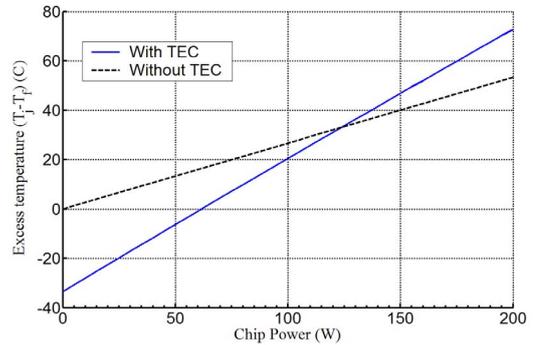

Figure 10. Power density limit for TEC.

The effectiveness of TEC varies under different external cooling conditions. Fig. 11 shows that if the external thermal resistance $\theta_{ext}$ is negligible, the use of TEC can lower the junction temperature $T_j$ by about 30°C. However, as the external thermal resistance increases, the benefit of using a TEC will gradually decrease. Therefore, the cooling benefit of using a TEC might be overestimated, if all the thermal resistances in Fig. 7 were ignored under the ideal assumption that the hot-side and cold-side temperatures are fixed [12]. It has also been shown in [13] that there is an optimum thickness of the TE legs and that the optimization of the current for maximum cooling should consider the heat sink thermal resistance.

The cooling rate can also be affected by the physical dimensions of TEC elements. Fig. 12 shows that the optimum leg length increases with thicker film. An optimum thin-film aspect ratio can therefore be derived. The existence of such an optimum aspect ratio results from the Joule heating effect and the Fourier heat transfer from the hot side to the cold side. It is to be noted that the results in Fig. 12 correspond to a fixed pitch of 10 um. If the TE element pitch were proportionally adjusted according to the film thickness, the excess temperature curves would collapse into one.

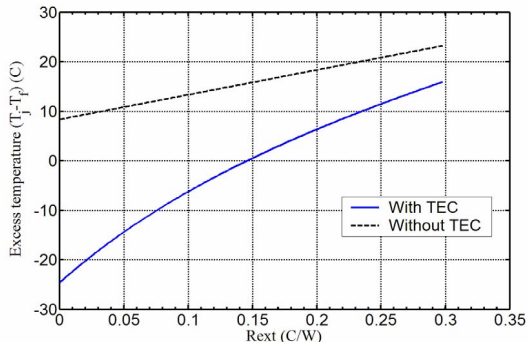

Figure 11. Effect of external cooling on TEC.

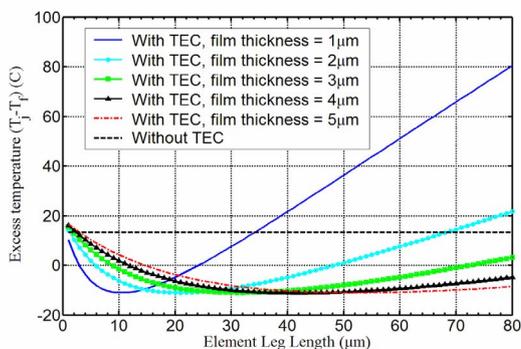

Figure 12. Effect of the TE element leg length.

Based on the analytical model and simulation results described above, the electric current and TE element dimensions can be optimized for thermoelectric cooling. Under these optimal conditions, sub-ambient cooling of the chip can be achieved provided that the power density is controlled within the target range. The parasitic thermal resistance components could limit the net cooling performance of TEC. The presence of other non-ideal factors such as thermal bypass effect due to the structural dielectric film, and the thermal and electrical contact resistance might also reduce the potential benefit of using TEC.

## 7. Conclusion

A new VLSI integration technology is proposed to achieve greater thermal efficiency with a high-density thin-film thermoelectric cooling module, where millions of thermoelectric couple elements are built on a semiconductor wafer. By using a mask-less doping technique, n-type elements are formed with a first angle implant using n-type dopants, and p-type elements are formed with a second angle implant using p-type dopants. Each pair of adjacent n-type and p-type thermoelectric couple elements is then connected by using a self-aligned silicidation process.

Comprehensive thermal simulations are conducted to optimize various TEC parameters for maximum heat removal. In addition, the thermoelectric cooling device enables precise temperature control through an integrated power monitoring and feedback system. The high-density thermoelectric arrays can be implemented directly on the backside of a semiconductor chip, embedded near the hot spots, or fabricated separately on a module. Combined with traditional heat sinks, micro channels, and heat pipes, the thermoelectric micro-coolers could significantly increase cooling capacity and provide a promising solution to the increasing power demand and thermal design challenges.

## References

[1] "Thermal management solutions for cooling electronics," http://www.aavidthermalloy.com .

[2] Scott Garner, "Heat pipes for electronics cooling applications," *Electronics Cooling*, vol. 2, no. 3, September 1996.

[3] Roger Schmidt, "Liquid cooling is back," *Electronics Cooling*, vol. 11, no. 3, August 2005.

[4] Ioan Saucic, "Thermoelectric & phase change technology building block with application to CPU cooling," Intel Technology Symposium, September 2004.

[5] S. Godfrey, "An introduction to thermoelectric coolers," Electronics Cooling, vol. 2, no. 3, pp. 30-33, Sept. 1996.

[6] Richard Chu, et al., "Electronic module with integrated programmable thermoelectric cooling assembly and method of fabrication," United States patent no. 6,548,894, April 15, 2003.

[7] J. Vandersande and J. Fleurial, "Thermal management of power electronics using thermoelectric coolers," in *Proc. 15th Int. Conf. Thermoelectrics*, 1996, pp. 252-255.

[8] Uttam S. Ghoshal, "Advanced thermoelectric cooling systems," Proceedings of Next-Generation Thermal Management Materials and Systems, June 2005.

[9] Clemens Lasance and Robert Simons, "Advances in high-performance cooling for electronics," Electronics Cooling, vol. 11, no. 4, November 2005.

[10] Robert Simons, Michael Ellsworth, and Richard Chu, "An assessment of module cooling enhancement with thermoelectric coolers," Journal of Heat Transfer, vol. 127, pp. 76-84, January 2005.

[11] Allan Kraus and Avram Bar-Cohen, *Thermal Analysis and Control of Electronic Equipment*, Hemisphere Publishing Corporation, New York, 1983.

[12] D.D.L. Wijngaards *et al.*, "Design and fabrication of on-chip integrated polySiGe and polySi Peltier devices," Sensors and Actuators, vol. 85, pp. 316-323, 2000.

[13] K. Fukutani and A. Shakouri, "Design of bulk thermoelectric modules for integrated circuit thermal management," *IEEE Trans. Components, Packaging and Manufacturing Technology*, vol. 29, no. 4, pp. 750-757, December 2006.